\documentclass[oldversion]{aa}

\usepackage{times,float}
\usepackage{epsfig}
\usepackage{graphics}
\usepackage{amssymb}
\usepackage{curves}
\usepackage{eepic}
\usepackage{epic}
\usepackage{txfonts}
\usepackage{natbib}

\newcommand{\swift} {{\it Swift}}
\newcommand{\simlt}{\,\hbox{\lower0.6ex\hbox{$\sim$}\llap{\raise0.2ex\hbox{$<$}}}\,}
\newcommand{\simgt}{\,\hbox{\lower0.6ex\hbox{$\sim$}\llap{\raise0.2ex\hbox{$>$}}}\,}
\newcommand{\fras}{\mbox{\ensuremath{.\mkern-4mu^{\rm s}}}}

\begin{document}

\title{The bright optical flash from GRB\,060117\thanks{Based on
observations of robotic telescope FRAM, operated by the Pierre Auger
collaboration at Los Leones site, Malarg\"{u}e, Argentina.}}
\titlerunning{The bright optical flash from GRB\,060117}

\author{ Martin Jel\'\i nek\inst{1} 
\and Michael Prouza\inst{2, 6} 
\and Petr Kub\'anek\inst{3, 4} 
\and Ren\'e Hudec\inst{3}  
\and Martin~Nekola\inst{3} 
\and Jan \v{R}\'{\i}dk\'y\inst{2, 6}
\and Ji\v{r}\'{\i} Grygar\inst{2, 6}
\and Martina Boh\'a\v{c}ov\'a\inst{2, 6}
\and Alberto J.~Castro-Tirado\inst{1} 
\and Javier~Gorosabel\inst{1} 
\and Miroslav Hrabovsk\'y\inst{2, 6}
\and Du\v{s}an Mand\'at\inst{2, 6}
\and Dalibor~Nosek\inst{5, 6}
\and Libor No\v{z}ka\inst{2, 6}
\and Miroslav Palatka\inst{2, 6}
\and Shashi B. Pandey\inst{1}
\and Miroslav Pech\inst{2, 6}
\and Petr Schov\'anek\inst{2, 6}
\and Radom\'{\i}r \resizebox{.5em}{.8em}{\v{S}}m\'{\i}da\inst{2, 6}
\and Petr~Tr\'avn\'{\i}\v{c}ek\inst{2, 6}
\and Antonio de Ugarte Postigo\inst{1}
\and Stanislav V\'{\i}tek\inst{1}
}
\institute{
Instituto de Astrof\'{\i}sica de Andaluc\'{\i}a (IAA-CSIC), Apartado de
Correos, 3.004, E-18.080 Granada, Spain.  
\and
Fyzik\'{a}ln\'{\i} \'{u}stav AV \v{C}R, Na Slovance 2, CZ-182 21 Praha
8, Czech Republic 
\and
Astronomick\'y \'ustav AV \v{C}R, Fri\v{c}ova 298, CZ-251 65
Ond\v{r}ejov, Czech Republic 
\and
{\it INTEGRAL} Science Data Centre, ch. d' Ecogia 16, CH-1290 Versoix,
Switzerland 
\and
\'{U}stav \v{c}\'asticov\'e a jadern\'e fyziky, MFF UK Praha, V
Hole\v{s}ovi\v{c}k\'ach 2, CZ-180 00 Praha 8, Czech Republic
\and
for the Pierre Auger Collaboration 
}

\abstract {We present a discovery and observation of an extraordinarily
bright prompt optical emission of the GRB\,060117 obtained by a
wide-field camera atop the robotic telescope FRAM of the Pierre Auger
Observatory from 2 to 10 minutes after the GRB.
We found rapid average temporal flux decay of $\alpha = -1.7 \pm 0.1$
and a peak brightness $R = 10.1$ mag.
Later observations by other instruments set a strong limit on the
optical and radio transient fluxes, unveiling an unexpectedly rapid
further decay. 
We present an interpretation featuring a relatively steep
electron-distribution parameter $p \simeq 3.0$ and providing a
straightforward solution for the overall fast decay of this optical
transient as a transition between reverse and forward shock.} 


\offprints{ \hbox{Martin Jel\'{\i}nek, e-mail:\,{\tt mates@iaa.es}}}

\date{Received  / Accepted }

\maketitle

\section{Introduction}

After the establishment of the GRB Coordinates Network (GCN)
\citep{bacodine} in 1995, the technical advances have enabled a fast and
reliable dissemination of satellite gamma-ray burst (GRB) data to
ground-based observers. 
Subsequently, wide use of the sophisticated robotic follow-up systems
has led to the first insight into the very early phases (less than 5
minutes after trigger) of the optical transients (OTs) accompanying some
GRBs. 
However, despite the extended efforts, optical data at the very early
stages are still sparse, so the whole picture remains unclear. 

The definition of the prompt optical emission was given by
\cite{piranold} as the optical emission arising during the
$\gamma$-emission period.
This early emission is usually explained in terms of either a reverse or
an internal shock \citep{saripiran99}.
With a certain delay after the GRB, the afterglow -- emission due to the
forward shocks -- starts to dominate the steeply decaying prompt
emission. 
This transition flattens the lightcurve.
The original rapid decay of the prompt emission slows down to a more
modest decay due to the afterglow.
It is generally accepted that the physical mechanisms of the prompt
emission and the afterglow are distinct. 
The distinction is probably not absolute, some observations suggest
\citep[cf.][]{chincarini} that at least the $X$-ray intensity of the
early emission phases can be extrapolated from the GRB emission itself. 

In this letter we present the observation of a very bright optical
transient associated with GRB\,060117 observed by the robotic telescope
FRAM. 

\begin{figure}[b!] 
        \begin{center} \label{fig1}
        \resizebox{\hsize}{!}{\includegraphics{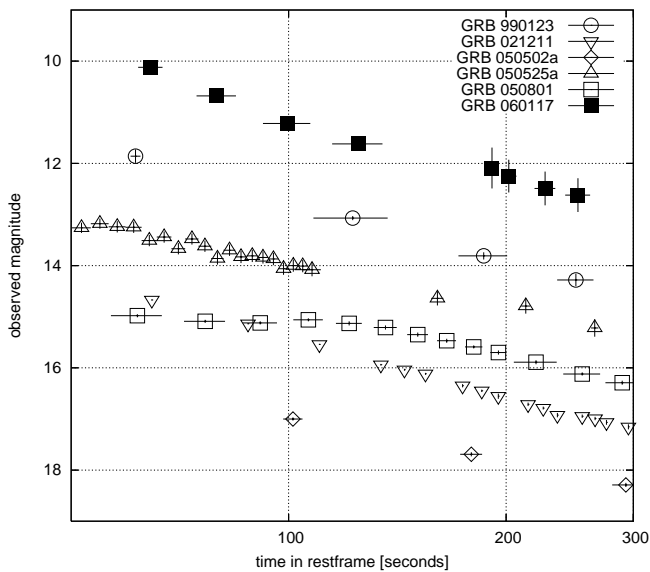}}
        \caption{The optical light curve of GRB\,060117 in comparison
        with other well-covered early GRB optical emissions: GRB\,990123
        \citep{Akerlof99}, GRB\,021211 \citep{Li03}, GRB\,050502a
        \citep{Guidorzi05}, GRB\,050525a \citep{Blustin06}, and
        GRB\,050801 \citep{Rykoff06}. 
        In this figure we use $z \approx 1.0$ for GRB 060117.  
        Observed $R$-band magnitudes are shown, except for GRB 050525a,
        where the $V$-band values are plotted.}
        \end{center} 
\end{figure}

\section{Observations}

A bright long-soft GRB\,060117 was detected by \swift\/ satellite on
January 17, 2006, at 6:50:01.6\,UT.
It showed a multi-peak structure with T$_\mathrm{90}$=$16\pm1$\,s with
maximum peak flux $48.9\pm1.6\,{\rm ph\,cm}^{\rm -2} {\rm s}^{\rm -1}$.
Coordinates computed by \swift\/ were available within 19\,s and
immediately distributed by GCN \citep{gcn4538}.

FRAM received the notice at 06:50:20.8\,UT, 19.2\,s after the trigger
and immediately started the slew. 
The first exposure started at 06:52:05.4, 123.8\,s after the GRB. 
Eight images with different exposures were taken before the observation
was terminated. 
A bright, rapidly decaying object was found, and its presence was
reported by \cite{gcn4535} and \cite{gcn4536} soon after the discovery.
The point-spread function of the object is similar to the stars in the
image, and the object did not move more than 2$''$ over the course of
our observation, ruling out a near-Earth object crossing the field of
view.
The weather conditions during the observation were very good, but the
Moon was nearly full (93\,\%) and the GRB location was only slightly
more than 5$^{\circ}$ above the horizon. 
Consequently, the magnitude limits of our observation were $\sim 2.5$
mag worse than the technical limit of the FRAM instrument in the optimal
conditions.
Table~1 displays the log of our observations (see also Fig\,1), where
the magnitude errors do not include systematic error of the USNO-B1.0 R1
magnitude, which should be $<$0.1\,mag.

An optical counterpart to GRB\,060117 was found 128.8\,s after the burst
at 
$$\alpha = 21^{\rm h}51^{\rm m}36\fras23  \quad  \delta=
-59^{\circ}58^{\prime}39\farcs3\quad \pm1\farcs5 \quad{\rm (J2000)}. $$
The error amounts to a 1-$\sigma$ uncertainty including systematic
errors. 

\section{Follow-up}

\swift\/ itself could not observe the GRB with its X-ray and optical
instruments, because of the Sun observing constraint \citep{gcn4533}.
One month later on Feb 14 and 15, 2006, \swift\/ XRT \citep{burrows05}
pointed to the burst position and did not detect any source at the
corresponding position with a 3-$\sigma$ limit of
1.0$\times10^{-3}$\,counts\,s$^{-1}$, corresponding to an unabsorbed 0.2
-- 10\,keV flux upper limit of
2.3$\times10^{-14}$\,erg\,cm$^{-2}$\,s$^{-1}$.
The burst was also detected by Konus-Wind \citep{gcn4542} and by Suzaku
WAM \citep{gcn4573}.

\begin{table} \begin{center} \label{table1}
        \caption{Optical R-band photometric observations of the optical
        flash GRB\,060117. 
        The magnitudes are not corrected for Galactic extinction ($A_{R}
        \sim 0.01$\,mag). $T - T_0$ is mean exposure time since the
        GRB.} 
\begin{tabular}{lcccccccc} \hline {\small UT Date of exp. start}&
{\small $T- T_0 [s]$} &{\small $T_{\rm exp}[s]$}& {\small $R$} &{\small
$\delta R$}
\\ \hline 
{\small 2006 Jan. 17.786169 }& {\small 128.8}& {\small 10}& {\small 10.12}& {\small 0.13}\\ 
{\small 2006 Jan. 17.786833 }& {\small 159.1}& {\small 20}& {\small 10.68}& {\small 0.12}\\ 
{\small 2006 Jan. 17.786343 }& {\small 199.3}& {\small 30}& {\small 11.22}& {\small 0.14}\\ 
{\small 2006 Jan. 17.787583 }& {\small 249.7}& {\small 40}& {\small 11.62}& {\small 0.18}\\ 
{\small 2006 Jan. 17.789109 }& {\small 382.4}& {\small 10}& {\small 12.09}& {\small 0.45}\\
{\small 2006 Jan. 17.789410 }& {\small 403.4}& {\small 20}& {\small 12.25}& {\small 0.36}\\
{\small 2006 Jan. 17.789815 }& {\small 452.9}& {\small 30}& {\small 12.49}& {\small 0.37}\\
{\small 2006 Jan. 17.790336 }& {\small 502.9}& {\small 40}& {\small 12.62}& {\small 0.37}\\
\hline 
\end{tabular} 

\end{center} \end{table}

Unfortunately, the later optical follow-up was unsuccessful due to
cloudy weather in both New Zealand and South Africa.
The limits reported by PROMPT \citep{gcn4548} (observations beginning
18.0 h after the burst), however, suggest a surprisingly rapid decay. 
The search for a radio afterglow was also unsuccessful \citep{gcn4547}.

The lag-luminosity pseudo-redshift estimation from \swift\/ data yields
$z \simeq 1.3 \pm 0.3$ \citep{gcn4538}. 
The redshift estimate based on the $\gamma$-ray data from Konus-Wind
gives $z \simeq 0.45\pm0.2$ \citep{gcn4544}. 

\begin{figure*}[t] 
        \begin{center} \label{fig2}
        \resizebox{\hsize}{!}{\includegraphics{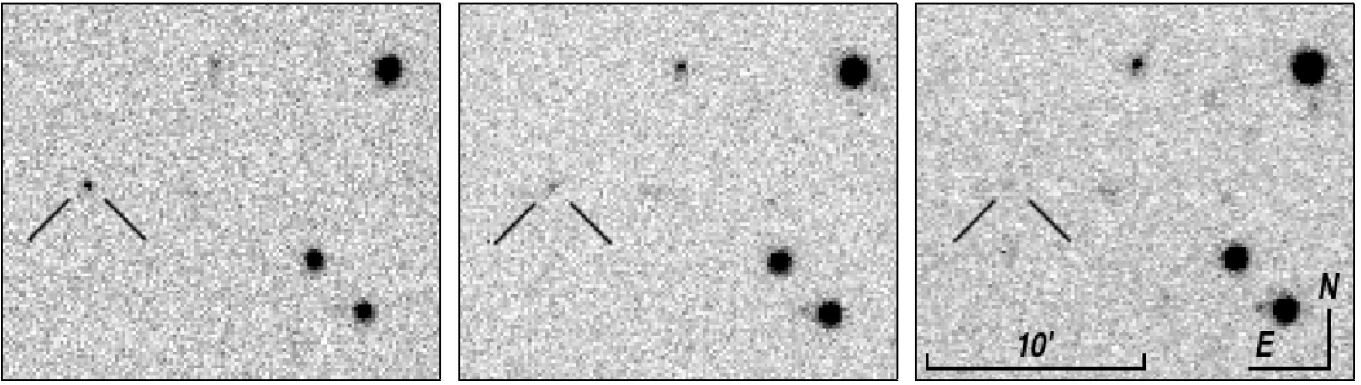}}
%
        \caption{Details of the surroundings of the optical flash of
        GRB\,060117 as observed by FRAM. 
        Images taken 129, 249, and 480s after the trigger.}
        \end{center} 
\end{figure*}

\section{Data acquisition and reduction}

FRAM is part of the Pierre Auger cosmic-ray observatory \citep{auger},
and its main purpose is to immediately monitor the atmospheric
transmission.  
FRAM works as an independent, RTS2-driven \citep{rts2}, fully robotic
system, and it performs a photometric calibration of the sky on various
UV-to-optical wavelengths using a 0.2\,m telescope and a photoelectric
photomultiplier. 
As a primary objective, FRAM observes a set of chosen standard stars and
a terrestrial light source. 
From these observations it obtains instant extinction coefficients and
the extinction wavelength dependence. 
%
Additionally, FRAM is able to follow GCN alerts, using its
wide-field camera with a fixed {\it R}-band filter. 

The wide-field camera consists of a Carl Zeiss Sonnar 200\,mm $f$/2.8
telephoto lens, SBIG ST7 imager, and Bessel R-band filter. 
The ST7 camera has a 768$\times$512 Kodak KAF-0402E CCD that covers a
field of view of $120^{\prime} \times 80^{\prime}$ with a scale of
$9\farcs6$/pixel.  
The effective diameter of the lens is 57\,mm and the $3 \sigma$ limiting
magnitude under optimum conditions reaches $R\sim15.0$ for a 30\,s
exposure.

The raw images were dark-frame subtracted using a median of several
dark-frame exposures. 
Given the significant dark current of the camera, the darks were treated
separately for each exposure time. 
The flat-field correction was then applied using the median of 40
normalized 1s exposures obtained while mosaicing through the twilight
sky.
The aperture photometry was done using the {\tt phot} routine in
IRAF\footnote{IRAF is distributed by the National Optical Astronomy
Observatory, which is operated by the Association of Universities for
Research in Astronomy, Inc., under co-operative agreement with the
National Science Foundation.} with the aperture diameter of 2 pixels.
To get a precise astrometric position of the source, we used the four
most significant images and computed the average position.

The images were astrometrically and photometrically calibrated on the
fly using the {\tt past} program in the context of JIBARO
\citep{jibaro}, using all sources with more than 10-$\sigma$
significance from the image compared to USNO-B1.0 \citep{usnob}
positions and {\it R1} magnitudes. 
{\tt Past} employs a sigma-clipped third-degree polynomial surface fit. 
For the astrometry, an error-weighted mean of the zero point is used.
Systematic errors of USNO-B1.0 should be less than 0\farcs2 in the
astrometry and less than 0.1\,mag for the photometry.
Since the Galactic extinction, taken from the maps published by
\cite{dust} is very low (E$_{\rm B-V} = 0.038$), we neglect this value
in the following discussion.

\section{Discussion}

In the search for the interpretation of the lightcurve, we assume a
uniform ISM, and that the influence of the internal shock emission on
the lightcurve is negligible because our observation starts $\sim$100\,s
after the end of the gamma-ray burst.

\begin{figure}[b!] 
        \begin{center} \label{fig4}
        \resizebox{0.95\hsize}{!}{\includegraphics{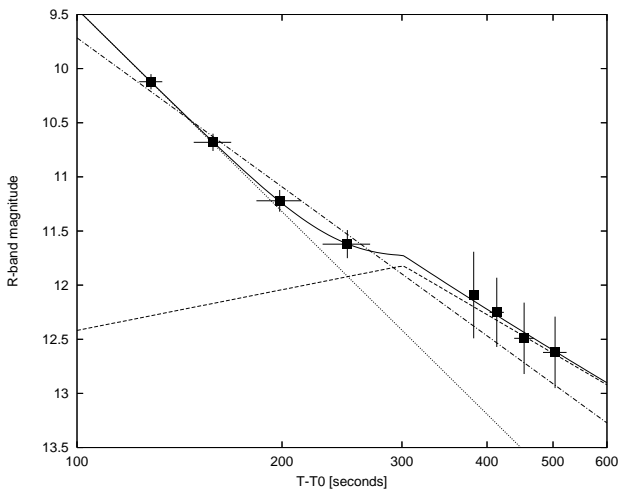}} 
        \caption{The R-band afterglow lightcurve of GRB\,060117. 
        The lightcurve is fitted as a superposition of reverse shock
        (dotted line) and forward shock (dashed line).
        The linear fit is plotted by a dot-and-dash line.}
        \end{center} 
\end{figure}

In the simplest case, the lightcurve can be fitted and a single power
law with a temporal flux-decay index $\alpha=-1.73\pm0.12$. 
The $\chi^2/d.o.f.$ for this fit is 1.296. 
If we assume this decay to be a signature of a pure forward shock, we
get the value of the electron energy distribution power-law index $p=3.3
\pm 0.1$. 
This value is very high in comparison with other known optical
transients, but it is consistent with $p=3.82_{-0.5}^{+1.0}$ as computed
from the Konus-Wind spectra of the GRB \citep{gcn4542}
\cite[cf.][]{shen}.
In contrast, if the linear decay is a signature of a pure reverse shock,
we get $p=2.0 \pm 0.1$ -- close to the typical value for the optical
transients observed so far. 
We should note that such a reverse shock emission should be accompanied
by a forward shock with $\alpha_{\mathrm F}\simeq$-0.7. 

Another possibility \citep[after][]{shao05} is to interpret the data as
a Type I lightcurve \citep[as given by][]{zhang03}, which depicts a
transition between the reverse and the forward shock with the passage of
the typical frequency break $\nu_m$ through the observed passband at
time $t_{m,f}$. 
We assume the lightcurve is initially dominated by a rapidly falling
reverse-shock emission with $F_{\nu,r}\sim t_d^{-(27p+7)/35}$, followed
by a forward-shock emission that rises as $F_{\nu,r}\sim t_d^{+1/2}$
before $t_{m,f}$ and then decays with $F_{\nu,f}\sim t_d^{-(3p-3)/4}$.
Using a $\chi^2$ minimization fit to this scenario, we obtain
$p=2.96\pm0.06$, a magnitude of forward shock maxima
$m_{m,f}=11.82\pm0.04$, a time of the maxima $t_{m,f}=301\pm4$\,s (after
trigger), and a magnitude of the reverse shock at $t=t_{m,f}$
$m_{m,r}=12.43\pm0.05$ ($\chi^2/d.o.f.$ for this fit is 0.015).
Corresponding decay indices are $\alpha_{\mathrm R}$=2.49$\pm$0.05 and
$\alpha_{\mathrm F}$=1.47$\pm$0.03 (see Fig\,3).
If the crossing time $t_\times$ \citep{zhang03} coincides with the end
of the GRB (i.e. $\sim$ 20\,s), then we can estimate the peak magnitude
of this OT as $R\sim 5$\,mag by backward extrapolation.

Note, that this is only one of the plausible interpretations. There may
be other possible explanations for this behaviour including density jump
in the media \citep{lazzati02} or energy injection \citep{bjornsson04}.  
Without a multiwavelength observation it is impossible to distinguish
which of these possibilities actually took place.  

The position of the burst and its distance from the Sun made the object
difficult to observe.
PROMPT \citep{gcn4548} shows that the bright OT decayed very fast, and
20 hours after the burst its magnitude was already $I > 21.2$. 
Using the procedure of \cite{simon} we transform this limit to the
filter of our observation: $R\simgt21.5$ 
From this limit we then get the estimate of an average late
decay as $\alpha_{\rm late} < -1.62$. 
Among the three interpretations we have shown, only the pure forward shock
scenario is compatible with this limit without introducing an unusually
early jet break, which is required to explain the other two scenarios.

Soon after the \swift\/ trigger, a suspicion of an extremely
low-redshift GRB was raised \citep{gcn4534} due to the presence of a
nearby galaxy PGC\,128172 with $z=0.04$ in the \swift\/ errorbox. 
However, this discovered transient lies 3\farcm1 from this galaxy,
accordingly the projected distance -- 160\,kpc -- is approximately four
times larger than the visible major diameter of this galaxy.
Furthermore, the position angle of the transient with respect to the
PGC\,128172, which we observe practically edge-on, is 97$^\circ$.
The association of the OT with this galaxy is, therefore, quite
unlikely.

\section{Conclusions}
\label{Conclusions}

The GRB\,060117 is the most intense (in terms of peak flux) GRB detected
so far by Swift. 
With the maximum brightness of $R = 10.1$ mag, FRAM has discovered one
of the optically brightest prompt optical emissions ever detected. 
The initial optical decay was found to be one of the steepest of an
early GRB optical afterglow observed. 

We have presented 3 scenarios for explaining the lightcurve. 
The apparent change in its slope is neglected in two simple scenarios,
where we suppose the observed lightcurve to only be the trace of a
forward, resp. reverse, shock. 
In the third (preferred) scenario, the shape of the lightcurve is
explained as a transition between reverse and forward shock emission.
The forward-shock-only interpretation is flawless regarding the PROMPT
limit, but shows rather spurious value of $p$. 
The other two scenarios (i.e. those involving reverse shock) result in a
relatively slowly decaying forward shock, and later limits require an
early jet break $t_{\mathrm j} \sim 0.2$\,d. The detailed analysis of
this problem is beyond the scope of this letter.

Progress in further study of the particular case of GRB 060117 depends
on the measurement of its distance.  
Therefore the follow-up and identification of host galaxy with a large
telescope is of high importance.  

A larger sample of GRB rapid follow-ups is needed to decide whether this
kind of transition, already suggested for other bursts, is common. 

\begin{acknowledgements}

The telescope FRAM was built and is operated under the support of the
Czech Ministry of Education, Youth, and Sports through its grant
programs LA134 and LC527.

MJ would like to thank to the Spanish Ministry of Education and Science
for the support via projects AP2003-1407, ESP2002-04124-C03-01 and
AYA2004-01515 (+ FEDER funds), MP was supported by the Grant Agency of
the Academy of Sciences of the Czech Republic grant B300100502.  
The currently used FRAM wide-field camera was obtained with support from
ESA PECS Project 98023.

We would also like to thank Primo Vitale for care for the telescope,
Rene Goerlich for his help with Gemini GOTO and Petr Heinzel, the
director of the Astronomical Institute of the Academy of Sciences of the
Czech Republic, for his generous support during the telescope testing at
the site of the Astronomical Institute in Ond\v{r}ejov.

\end{acknowledgements}


\hyphenation{Post-Script Sprin-ger}

\end{document}